\documentclass[graybox]{svmult}

\usepackage{mathptmx}       % selects Times Roman as basic font
\usepackage{helvet}         % selects Helvetica as sans-serif font
\usepackage{courier}        % selects Courier as typewriter font

\usepackage{makeidx}         % allows index generation
\usepackage{graphicx}        % standard LaTeX graphics tool
\usepackage{multicol}        % used for the two-column index

\makeindex             % used for the subject index
\usepackage{color}
\usepackage{amsfonts}
\usepackage{amsmath,amssymb,amsfonts}
\usepackage{mathrsfs}
\usepackage{graphicx}
\usepackage{natbib}
\usepackage{bibentry}
\usepackage{setspace}
\usepackage{tikz}

\numberwithin{equation}{section}
\newtheorem{Theorem}{Theorem}[section]

\newtheorem{Remark}{Remark}[section]

\newtheorem{Definition}{Definition}[section]
\newtheorem{Algorithm}{Algorithm}[section]

\newcommand{\EE}{{\mathbb E}}

\newcommand{\PP}{{\mathbb P}}

\newcommand{\xin}{\xi^N}
\newcommand{\hxin}{\hat{\xi}^N}
\newcommand{\hxi}{\hat{\xi}}
\newcommand{\F}{\mathcal{F}}
\newcommand{\R}{\mathbb{R}}
\newcommand{\rp}{R^{\mathrm{loc}}_t(\xin,\xi)}
\newcommand{\rph}{R^{\mathrm{loc}}_t(\hxin,\hxi)}
\newcommand{\cov}{\mathop{\mathrm{cov}}}

\newcommand{\argmin}{\mathop{\mathrm{arg min}}}

\usepackage{color}

\newcommand{\VM}{\mathcal{V}}
\newcommand{\NPV}{Intrinsic Value }
\newcommand{\NPVs}{Intrinsic Values }

\begin{document}

\title*{A Hedged Monte Carlo Approach to Real Option Pricing}
\author{Edgardo Brigatti, Felipe Macias, Max O.  Souza, and Jorge P. Zubelli}
\institute{Edgardo Brigatti \at IF, UFRJ, 
Av. A. da Silveira Ramos, 149,
% Cidade Universit\'{a}ria, 
Rio de Janeiro, RJ 21941-972, Brazil. \email{edgardo@if.ufrj.br}
% \and Felipe Macias \at IMPA,
% Est. D. Castorina 110, Rio de Janeiro, RJ 22460-320, Brazil. \email{fmacias@impa.br},
\and Max O.  Souza \at DMA, UFF, Brazil.
 R. M. S. Braga, s/n, Niteroi, RJ 22240-120, Brazil. \email{msouza@mat.uff.br},
\and Felipe Macias \& Jorge P. Zubelli \at IMPA,
Est. D. Castorina 110, Rio de Janeiro, RJ 22460-320, Brazil. \email{fmacias,zubelli@impa.br}
}
\maketitle

\abstract{
In this work we are concerned with valuing optionalities associated to invest or to delay investment  
%the option to invest 
in a project when the available
information provided to the manager comes from simulated data of cash flows under historical
(or subjective) measure in a possibly incomplete market. Our approach is suitable also to incorporating 
subjective views from management or market experts and to stochastic investment costs. \\
It is based on the Hedged Monte Carlo strategy proposed by \cite{potters2001hedged}
where options are priced simultaneously with the determination of the corresponding hedging.
The approach is particularly well-suited to the evaluation of commodity related projects whereby the availability of
pricing formulae is very rare, the scenario simulations are usually available only in the historical measure, and the cash flows 
can be highly nonlinear functions of the prices.   
}

\section{Introduction}

The use of quantitative finance techniques to evaluate projects while trying to capture the value of active management and flexibility is 
known by the name of \textit{Real Option Analysis} (ROA). The importance of capturing such ``non-passive" value of projects can be a 
decisive factor when trying to decide upon investment within  a portfolio of projects.  Most of the classical applications of ROA 
involves vanilla American options as the case of the option to postpone a project, or to abandon it. However, when considering projects
related to capacity planning, chemical or petrochemical plants,  oil refining, 
or indeed any commodities-based project, a significant increase in complexity arises.  Under these conditions, recurring problems that are encountered in real options, 
as dealing with market incompleteness,  become particularly acute. 

In many cases, the company has access to  financial instruments that strongly correlate with the projects, and sometimes,  
as in the case of commodities companies, even with their final product.  Thus, the company can hedge some of its exposition yielded by a project, but usually not all of it, by an appropriate hedging portfolio.  This suggests that a hedging approach based on Monte Carlo simulations  can be  a plausible alternative for pricing such real options.  Indeed, on one hand  quadratic hedging has been used to price financial options in incomplete markets, and it is based on the local minimization of a proxy to variance, that is readily recognized as a risk measure by managers.  On the other hand, Monte Carlo approach has been often used when dealing both with options involving many assets---as baskets, rainbow, etc.---or when asset price models are not readily available. 

The aim of this work is to propose the use of the so-called Hedged Monte Carlo Method---Monte Carlo pricing through quadratic hedging---to price such complex options.

% \subsection{Outline}

% \marginpar{Placed outline here}
The plan for this article goes as follows: We close this introductory section with a 
description of the project evaluation problem we are considering, a 
short methodological review of the different
approaches to real options,  and its analysis by means of hedging with financial  instruments. 
% In Section~\ref{sec:metrev} we present a short methodology review and some further 
% background on the different approaches to Real Options.  
In Section~\ref{TheHMC} we present an approach to evaluating real options based on the Hedged Monte Carlo (HMC) method of \cite{potters2001hedged}. 

It has a number of desirable features: it uses the dynamics under the historical/subjective measure; 
it allows for an easy determination of the optimal exercise boundary,  it has low variance,
and allows for an assessment of the nonhedgeable risk. Furthermore, 
the oracle approach easily allows to incorporate managerial views in many different levels: 
it can either accommodate views of different managers of related projects, or more global corporative views and scenarios. 
% Furthermore, the implementation is quite straight-forward and one can easily incorporate managerial views. 
The method developed is explored in Section~\ref{examples} with some examples and a few case studies. 
We conclude in Section~\ref{sec:conclude} with some final comments and suggestions for further developments.

\subsection{Real options analysis}

The use of mathematical finance techniques has been continuously growing in recent times as a 
tool to capture the value of flexibility  in projects. A classical account can be found in the books of \cite{dixitpindyck94}
and  \cite{Trigeorgis1999}. 
The subject blossomed under different names but is generally known {\em Real Options}. See also  
\cite{MYERS1977,Touri1979,TITMAN1985,BrSc85,MCDONALD1986,TG1987,PADDOCK1988,DIXIT1989,PINDYCK1991,INGERSOLL1992}.

The original framework  identifies the Net Present Value of the project as a stochastic process correlated with a tradable risky asset.  
The risky asset  is termed the {\em twin} or {\em spanning} asset whereas the project value is sometimes referred as the {\em surrogate}
asset. 
Subsequent approaches take this identification very far.  Indeed, one cannot expect to have a traded asset with a perfect correlation 
with the project, since this would mean that project risk is totally diversifiable, and hence replicable via financial markets.
An alternative view, is to look for an asset, typically an index, that yields a high correlation with the project returns. 
This is known as the {\em modern approach}. Other approaches exist. See \cite{Borison2005} for a classification, the 
discussion in \cite{Jaimungal2011}, and the remarks in Section~\ref{sec:metrev}. 

A very strong critique of the real option approach was presented by \cite{HS2001}. There they show, by means of a simple
example, that the use of no-arbitrage techniques to nontradable surrogate assets can lead to arbitrary (very high or very low) no-arbitrage
option prices. This in turn shows that the economic use of real options in the context of incomplete markets is highly questionable. 
In the same work, they also show that a variance minimization of the hedging error could be a way out of the economical impasse
caused by the lack of completeness of the market. 

\subsection{Complex structured  real options}

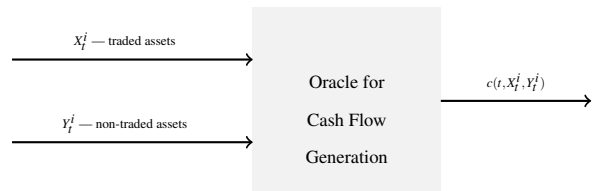
\begin{figure}[h!]
\begin{center}
\begin{tikzpicture}
% \draw [thick] (4,4) -- (4,0) -- (0,0) -- (0,4) -- (4,4);
\fill [gray!10,thick,->] (0,0) rectangle (2.5,2.5);
\draw [thick,->] (-3.2,1.8) -- (0,1.8);
\draw [thick,->] (-3.2,0.7) -- (0,0.7);
\draw (-1.7,1.8) node [above] {{\tiny $X_t^i$ --- traded assets}};
\draw (-1.7,0.7) node [above] {{\tiny $Y_t^i$ --- non-traded assets}};
\draw (1.25,1.5) node {{\scriptsize Oracle for}};
\draw (1.25,1) node {{\scriptsize Cash Flow}};
\draw (1.25,0.5) node {{\scriptsize Generation}};
\draw [thick,->] (2.5,1.25)--(4.5,1.25);
\draw (3.5,1.25) node [above] {{\tiny $c(t,X_t^i,Y_t^i)$}};
\end{tikzpicture}
\end{center}
\caption{ Description of the oracle producing the cash flows at time $t$ and scenario $i$. \label{pic1}}
\end{figure}

We are concerned with  the practical problem of quantitatively evaluating projects under uncertainty from different scenarios 
taking into account flexibility of the projects and the possibility of partial hedging with financial instruments. 
We assume that we have available  a fairly large number
% ~\footnote{How large the 
% number of scenarios must be is the subject of a discussion that goes beyond the scope of this article.} 
of scenarios organized in a time series and that connected to the different scenarios we have an {\em
oracle} that produces the cash flows associated to each scenario. The scenarios in turn are linked to traded assets or financial instruments 
which may be used for hedging the project. Figure~\ref{pic1} describes the situation. 

% \bigskip
This framework can arise  when planning chemical plants or oil refineries. See for example \cite{MZP98}, \cite{P2009}, \cite{SGFC1989}, 
\cite{S2009},  \cite{Oetal2007}. It also naturally appears when using real option techniques  
for capacity planning. See \cite{M2004}, \cite{NS2005}, \cite{CCC2007}.  
In most of these problems, the markets are overall incomplete, unless under very simplifying assumptions. 
In addition, such incompleteness will  also imply that data will be only available under the historical measure. 

We shall now consider different ingredients in such complex options. The first one, stems from the fact that 
many corporations predict in a fairly precise way their cash flows using a black box (oracle) whose stochasticity only comes through
the inputs from the different assets, supply/demand curves, and production curves. Yet, such oracle
depends on the prices of many (stochastic) assets as well as on non-tradable quantities. This is depicted in the Figure~\ref{pic1}.

More generally, the cash flows may be produced by simplified models that incorporate algorithms or analytical procedures.

% 
% {\paragraph{Challenges in using Real Options}}
Among the challenges that are present in the evaluation of projects under uncertainty, especially those linked to 
commodity enterprises, we single out the following: 
\begin{itemize}
\item {\em Historical measure:} The simulations are usually presented in the historical measure. Furthermore, the scenarios are provided by management and
are loaded with views from advisors or sector specialists. In fact, some corporations delegate the scenario generation to part of the
board of directors or an independent division. 
% \changes{Thus, we do not have access easy to risk-neutral simulations.}
% {Thus we only have access to historical simulations.} % redundant!
\item {\em Managerial views:} It is crucial to incorporate managerial views in the cash flows, as well as automated decisions. An example would be
a commodity trading company that has a limited amount of storage capacity for different products. According to the relative
prices and profits it may automatically determine how much of each product it would store. 
\item {\em Market incompleteness:} The hedging is performed in incomplete financial markets. In fact, sometimes the firm does not
have access to the liquidity provided by the financial markets. In other cases, regulations might preclude the management to hold 
some speculative positions to fully hedge against market variations. 
\item {\em Unhedgeable risks:} Investment decisions on commodity related projects have to take into account not only the hedgeable
risks, but also the unhedgeable ones. For instance, the decision of exploring an oil field is highly dependent on its production curve
and also on ecological risks associated to the operation. 
% Pricing and hedging of Real Options in incomplete financial markets
%How to price options with Monte Carlo methods if some of the variables are not traded? 
\item {\em Multiple assets:} Investment decisions may depend on the relative value of several traded underlyings. Such assets
might have general correlation structures ranging from low to high cross-correlation. Thus, the hedging might have to be 
very diversified.
% Handle many multiple assets (corporations trade on a large number of assets in different markets)

\end{itemize}

\subsection{Real option analysis through hedging}

The approach suggested here to attack the general problem mentioned above can be loosely described as a risk minimization one where the project
valuation is performed by constructing a portfolio that includes the project delay optionality and the possible hedging of such project by
tradable assets. By a methodology introduced by \cite{potters2001hedged} (see also %Grasselli and Hurd
\cite{gh2004}) one can compute different financial 
options (including American and Bermudian ones) by a recursive risk minimization of {\em historical-measure} simulated paths. 
The importance of using historical simulations in the solution of this problem is that managers consider their
decisions by looking at observed prices of different commodities and assets. We shall refer to the methodology developed 
in \cite{potters2001hedged} by the {\em Hedged Monte Carlo method} (HMC).

Another motivation for the methodology presented here is the critique to the traditional no-arbitrage
arguments of real option theory present in the work of \cite{HS2001}. In the latter, the idea of minimizing the variance is
considered as an alternative to the shortcomings caused by market incompleteness. 
A number of different approaches have been developed to deal with incomplete markets. To cite a few: 
indifference pricing, minimal martingale measure, and minimal entropy measure. 
% 
% 
% paper we favor the approach of minimal martingale measures of \cite{SF1988}. Such approach can be implemented in 
% practice by HMC of \cite{potters2001hedged} as 

The idea of using HMC or Monte Carlo algorithms to compute option prices in incomplete markets is not new. See, for example,
the work of \cite{PY2008} and the references therein. It can also be traced to the preprint of \cite{gh2004}. The novelty
of the approach suggested herein is the idea of incorporating the different cash flows in the evaluation, producing the 
different statistics that may be helpful for the manager and allow for the possibility of incorporating managerial views in the
simulations. 
As it turns out, the HMC methodology corresponds to choosing the minimal martingale measure of \cite{SF1988}. 
See \cite{Lipp2012} and  \cite{Gastel2013} and references therein for details on such connection.

\subsection{Remarks on alternative approaches}\label{sec:metrev}

We shall now  briefly review the various methodologies available to price real options.

{\bf Hedging Public and Private Risks:} 
As observed in the works of \cite{Borison2005} and of \cite{Jaimungal2011}, one of the main issues  in evaluating  
different types of projects is whether the source of risk is public or private. 
For projects with returns that are highly correlated to the market,  risk mitigation  should be almost completely achievable by hedging 
it  with traded assets.  In most approaches, the project is assumed to be perfectly correlated to  a single asset, and hence replicable. 
Notice that for projects which have a diverse range of products, it might be necessary to use a basket of hedging assets.  

 On the other hand,
projects with mainly private risks, such as for instance R\&D, are unlikely to be hedged   with  the use of traded assets. 
Moreover, in some  cases the 
valuation of the project can be highly dependent on management estimates.  Thus, one can think of  such estimates as a non-traded asset 
that contributes to the value of the project. 

From the point of view of utility theory, this can be more precisely measured by specifying the firm's preferences through a 
utility function, and thus one can think of using indifference pricing. This approach was pursued in a number of works, in 
particularly in the work of \cite{HH2002}, of \cite{gh2004}, and of \cite{mr2349306}.

{\bf The Classical Method: }
As mentioned in the introduction, the classical methodology of pricing real options assumes that there 
is a spanning asset that is highly correlated to the net present value (NPV)
of the project. One example of such methodology is the so-called  
% \paragraph
{\em Marketed Asset Disclaimer (MAD) Approach} is based on the idea of taking the  NPV 
distribution both as the value of the project and as the underlying (tradable) asset. Then, model the asset with a stochastic dynamics
and perform Risk-Neutral pricing, perhaps accounting for non-traded issues. See for example \cite{Copeland2001} and 
\cite{copeland2004real}. 
% \begin{enumerate}
% \item Take the NPV distribution both as the value of the project and as the underlying asset.
% \item Model the asset with a stochastic dynamics
% \item Perform Risk-Neutral pricing, perhaps accounting for non-traded issues.
% \end{enumerate}
Among the advantages we mention that it mimics the standard mathematical finance approach, the theory is 
fairly simple and many out-of-the-box numerical methods are available.
% 
% {\bf Advantages}
% \begin{enumerate}
% \item Very close to classical view of Real Options and to standard mathematical finance approach.
% \item Simple theory and numerical methods are available.
% \end{enumerate}
%
As for the disadvantages, besides the general criticism mentioned before in reference to the work of \cite{HS2001}, we should also 
note that often very few data is available for calibration. This makes the choice of the underlying dynamics somewhat arbitrary.
Furthermore, for each project, a calibration/choice of underlying dynamics is necessary. 
This ambiguity is typically tackled by a simplifying assumption on the dynamics, 
 which will hopefully be consistent with the market scenarios.
% This might not be a trivial task.
% {\bf  Disadvantages}
% \begin{enumerate}
% \item Very few data available for calibration. This makes the choice of the underlying dynamics somewhat arbitrary.
% \item For each project, a calibration/choice of underlying dynamics is necessary. This might not be a trivial task.
% \end{enumerate}

{\bf Monte Carlo Based Approaches:} 
% 
% % {\bf The Monte Carlo Approach (Based on \cite{LongstaffSchwartz})}
% 
% % This was the approach taken by Lillywhite. 
In many situations the project or the firm has a simulator that we shall refer from now on as an {\em oracle}. 
Such oracle produces information about the cash flows associated to different projects or optionalities for different
scenarios which in turn are generated from inputs of tradable assets. 
The idea is then to take the oracle output as the payoff distribution, and use the method of \cite{Longstaff2001}
to compute the corresponding conditional expected values subject to the traded asset prices. This requires the
underlying(s) to be simulated in the risk-neutral measure or taking into account the market price of risk in the
final result. 

% \begin{enumerate}
%  \item Take the oracle as a payoff distribution.
% \item Use the traded inputs for oracle together with  Longstaff-Schwartz to evaluate option pricing
% \item Pricing is made on the Physical measure. Market price of risk is then interpreted as the dividend of the project.
% \item This approach is compatible with a CAPM view of the world. 
% \item Alternatively, one could run oracle simulations with risk-neutral versions of the traded assets.
% \end{enumerate}

% \end{frame} 
% \begin{frame}

%{The Monte Carlo Approach (Based on \cite{LongstaffSchwartz})}{(Cont.)}
% {\bf  Advantages}
% 
% \begin{enumerate}
% \item Uses fully the oracle information towards the option evaluation.
% \item Easily integrated within oracle, with project independent pricing evaluation.
% \item Great managerial appeal.
% \end{enumerate}
Among the pros of such approach, we should mention that it uses fully the oracle information towards the option evaluation, 
it is easily integrated and automated with the  oracle thus leading to a  project independent pricing mechanism. Furthermore, it has
a good managerial appeal. As for the cons, we have that since the simulation is performed on the oracle data, 
the realizations are restricted to the ones
generated by the oracle. This can impair the quality of the results obtained.
Furthermore, the risk-neutral calibration of the scenario generation that will provide inputs to 
the oracle might be very cumbersome and requires extra work. 

% \end{frame} 
% \begin{frame}
% 
% 
% {\bf  Disadvantages}
% \begin{enumerate}
%  \item Since the simulation is made on the oracle data, the realizations are restricted to the ones
% generated by the oracle. This can be impair the quality of the results obtained.
% \item Although preliminary studies with simulations using Physical measures and Risk-Neutral do not give significant differences, there is lack of theory for justification.
% \end{enumerate}
% \end{frame}
% \section{Other Approaches in the Literature}
% 
{\bf Datar-Mathews (DM) Method: }
In the method proposed by \cite{dm1}
one assumes that it is given the NPV distributions (usually by management).
% This is identified with the payoff, and the identify these payoffs as the value of the project.
Then, one performs a Monte Carlo simulation to replicate the distribution at the given times and to produce 
a simulated process for the underlying asset. %\todo{this has to be improved}
% \item Use the Monte Carlo simulated process as an underlying asset and pricing accordingly.
% \item Notice that discounting is done with different rates: strikes by the risk-free rate, underly 
% by the corporate bond rate. Again, this can be made consistent with CAPM view of the world.
% 
% \begin{enumerate}
%  \item Assume one is given NPV distributions; usually managerial given.
% \item Identify these as payoff, and the identify these payoffs as the value of the project.
% \item Perform a Monte Carlo simulation to replicate the distribution at the given times 
% \item Use the Monte Carlo simulated process as an underlying asset and pricing accordingly.
% \item Notice that discounting is done with different rates: strikes by the risk-free rate, underly by the corporate bond rate. Again, this can be made consistent with CAPM view of the world.
% \end{enumerate}

Among the advantages, we can mention that it is easily implemented and has great managerial appeal.
Yet, there is lack of theory to justify such approach. 

% % \end{frame} \begin{frame}
% {\bf  Advantages}
% \begin{enumerate}
%  \item  Easily implemented. 
% \item Great managerial appeal.
% \item Integrates ROA with present managerial methods smoothly.
% \end{enumerate}
% 
% {\bf Disadvantages}
% \begin{enumerate}
%  \item There is a lack of theory to justify this approach.
% \end{enumerate}

% \end{frame} 

% \begin{frame}

{\bf Jaimungal-Lawryshyn (JL): }
The work of \cite{Jaimungal2011} includes an extension of DM method as follows: They take the NPV distributions and 
choose an observable sector index (not-traded on their paper) that is highly correlated with cash flows. They 
choose a dynamics for this index and 
based on the dynamics, find the payoff functions that yield the NPV distribution as a function of this market index.
Then, they identify the value of the project as expected values of these payoffs (very much line in DM's method). 
Finally, they 
choose a correlated (if possible) traded asset or index and perform a risk-neutral valuation using a Minimal Martingale Measure. 

Among the advantages of this method, we can cite that as in the DM method, it integrates the managerial view with the 
Real Option Analysis. Thus it has a good managerial appeal. Furthermore, the theory is more sound. Yet, the market index might not be 
easily available and one still needs to calibrate the model to the index. This step might be hard if the data is not abundant.

\section{The Hedged Monte Carlo Approach and Minimal Martingale Measures}
\label{TheHMC}

Since the typical data that will be used for the method comes from simulations, %or from point wise estimation of values, 
it will  be naturally discrete in time. Thus, it is natural to adopt a discrete time approach for the algorithm. 
In this vein, we begin by reviewing the theory for quadratic hedging in discrete time and how it can be used to price contingent claims. 
This will follow closely the exposition of \cite{Follmer:Schied:2004}.
Then, we proceed on to discussing the algorithm itself, and present a brief remark about its relation to a 
continuous version of the problem.

\subsection{Hedging in discrete time  within an incomplete market: a review}
In an incomplete market setting, from its very definition, a self-financing replicating strategy is not usually available. In this scenario, one might  give up the replicating property, and look for self-financing hedging strategies that control  the down side risk---evaluated by 
means of a risk measure. See for example the work of \cite{Follmer:Schied:2004}. Alternatively,  one enforces a replicating strategy and 
looks for the cheapest strategy with this property. 
In this latter case, a very popular strategy among practitioners is  the  minimization of the quadratic tracking error  \cite{Schw2008}.
% THE REFERENCE IS: 
% M. Schweizer (2008) 
% Local Risk-Minimization for Multidimensional Assets and Payment Streams
% Banach Center Publications 83, 213-229 
This choice leads to strategies that are self-financing in the mean under very mild assumptions, that we now briefly review.

As usual, we assume to be in a filtered probability space $(\Omega,\F_T,\PP)$ and write $L^2(\PP)=L^2(\Omega,\F_T,\PP)$,
where $\PP$ denotes the historical measure.  
In what follows, $\xin$ denotes the investment (short or long) in the num\'eraire asset, and $\xi$ denotes the position on 
$d$ risk assets, with prices given by a $d$-dimensional stochastic process $X$. Furthermore, $X$ and $V$ denote {\em
discounted prices} with respect to a risk-free process.
\begin{Definition}
%start mod by fm 
A trading strategy is a pair of stochastic processes $(\xin,\xi)$, where $\xin_t$ is an adapted process   
and $\xi$ is a $d$-dimensional predictable process.
The discounted value of the portfolio is 
\[
V_t:=\xin_t+\xi_t\cdot X_t
\]
The gain process is 
\[
G_t:=\sum_{s=1}^t\xi_s\cdot \left(X_s-X_{s-1}\right).
\]
The cost process is defined as
\[
C_t:=V_t-G_t.
\]
%end mod by fm 
\end{Definition}

Let $H$ denote a random claim, and assume that
\begin{enumerate}
\item $H \in L^2(\PP)$;
\item $X_t \in L^2(\Omega,\F_T,\PP;\R^d)$, for all $t$.
\end{enumerate}
\begin{Definition}
An admissible $L^2$-strategy for $H$ is a trading strategy such that it is replicating, i.e., 
\[
V_T=H\quad \PP\,\text{ a.s.}, \quad  
\]
and such that both the value process and the gain process are square-integrable, i.e., 
\[
V_t,G_t\in L^2 (\PP),\,\forall t\in [0,T].
\]
\end{Definition}
We can now introduce a suitable risk process
\begin{Definition}
Let $(\xin,\xi)$ be an $L^2$-admissible strategy. The corresponding local risk process is given by
\[
\rp=\EE[(C_{t+1}-C_t)^2|\F_t].
\]
Let  $(\hxin,\hxi)$ be an $L^2$-admissible strategy with value process $\hat{V}_t$. This strategy is  said to be a locally risk-minimizing strategy if, for each $t$, we have that
\[
\rph\leq \rp,\quad \PP\,\text{a.s.}
\]
for each $L^2$-admissible strategy whose value process $V_t$ satisfies $V_{t+1}=\hat{V}_{t+1}$.
\end{Definition}
\begin{Definition}
A trading strategy is a mean self-financing strategy, if its corresponding cost process is a martingale, i.e.:
\[
\EE[C_{t+1}-C_t|\F_t]=0.
\]
\end{Definition}
\begin{Definition}
We say that two adapted processes $U$ and $V$ are strongly orthogonal if 
\[
\cov(U_{t+1}-U_t,V_{t+1}-V_t|\F_t)=0,
\]
where $\cov$ denotes the conditional covariance, i.e., 
$\cov(A,B|\F_t)= \EE[AB |\F_t] - \EE[A |\F_t]\EE[B |\F_t]$.
\end{Definition}
The following result (see \cite{Follmer:Schied:2004}) guarantees the existence of the corresponding hedge: 
%JPZCHANGE The main result that will show the existence of the corresponding hedge:
%
\begin{Theorem}
\label{thm:main}
\begin{enumerate}
\item An $L^2$-admissible strategy is locally  risk minimizing if, and only if, it is mean self-financing, and its cost process is strongly orthogonal do $X$.
\item There exists a locally risk minimizing strategy if, and only if, $H$ admits the so-called Follmer-Schweiser decomposition:
\[
H=c + \sum_{t=1}^T\xi_t\cdot (X_t-X_{t-1} )+L_T,\quad \PP\,\text{-a.s.},
\]
where $c$ is a constant, $\xi$ is a $d$-dimensional predictable process, such that $\xi_t\cdot(X_t-X_{t-1})\in L^2(\PP)$ for each $t$, 
and $L$ is a square-integrable martingale that is strongly orthogonal to $X$, and satisfies $L_0=0$. 

In this case, the locally risk-minimizing strategy $(\hxin,\hxi)$ is given by : %$\hxi=\xi$, and
\begin{align*}
\hxi&=\xi\\
\hxin_t&=c + \sum_{s=1}^t\xi_s\cdot (X_s-X_{s-1}) +L_t-\xi_t\cdot X_t.
\end{align*}
Notice that the associated cost process is $C_t=c+L_t$.
\end{enumerate}
\end{Theorem}

\subsection{Pricing by risk minimization}
The proof of Theorem~\ref{thm:main} is actually constructive and yields the following algorithm:
\begin{Algorithm}\label{gen:algo}
\ 
\begin{enumerate}
\item  Set $\hat{V}_T:=H$;
\item For $t=T-1$ down to $t=0$ do %%%%%%%%%%%%%%%%%%% ATTENTION JPZCHANGE LATE AT NIGHT TO t=0
\begin{enumerate}
\item Set 
\begin{equation}\label{variation}
% begin mod fm
\displaystyle (\hat{V}_t,\hxi_{t+1}):=
\argmin_{(V_t,\xi_{t+1})} \EE\left[\left(\widehat{V}_{t+1}-(V_t+\xi_{t+1}\cdot(X_{t+1}-X_t))\right)^2\big|\F_t\right]
;
% end mod fm
\end{equation}

\end{enumerate}
\item Set $\hat{C}_t:=\hat{V}_t-\sum_{s=1}^t\hxi_s\cdot(X_s-X_{s-1})$, $t=0,\cdots,T$;
\item Set $\hat{c}:=\hat{C}_0$;
\item Set $\hat{L}_t:=\hat{C}_t-\hat{c}$, $t=0,\cdots,T$;
\item Set ${\hxin}_t:=\hat{c} + \sum_{s=1}^t\hat{\xi}_s\cdot (X_s-X_{s-1}) +\hat{L}_t-\hat{\xi}_t\cdot X_t$, $t=0,\cdots,T$.
\end{enumerate}
\end{Algorithm}

Notice that if $\PP$ is a risk-neutral measure, then $X_t$ is a square-integrable  martingale. In this case,  
the {G}altchouk-{K}unita-{W}atanabe decomposition (\cite{CS96Watanabe}) yields
\[
\EE[H|\F_t]=\hat{V}_0 + \sum_{s=1}^t\hat{\xi}_s\cdot (X_s-X_{s-1})+L_t
\]
and hence we have
\[
\EE[H|\F_t]=\hat{V}_t.
\]
This allows for  a consistent interpretation of the value of a local risk  minimizing strategy as an arbitrage-free price of $H$.
However, in general, $X$ will not be a martingale under $\PP$, and in the incomplete setting there will be many martingale measures that are equivalent to $\PP$. It turns out that one of these measures is  particularly relevant for  hedging under local risk minimization. 
\begin{Definition}
Let $\mathcal{P}$ denote the set of martingale measures that are equivalent to $\PP$. We say that $\hat{\PP}\in\mathcal{P}$ is a minimal martingale measure if
\[
\EE\left[\left(\frac{d\hat{\PP}}{d\PP}\right)^2\right]<\infty,
\]
and if  every square-integrable martingale under $\PP$, which is strongly orthogonal to $X$ is also a martingale under $\hat{\PP}$.
\end{Definition}
\begin{Theorem}
\label{thm:eqv_mm}
If there exists a minimal martingale measure $\hat{\PP}$, and denoting by $\hat{V}$ the value process of 
the local risk minimizing strategy, then we have that
\[
\hat{V}_t=\hat{\EE}\left[H|\F_t\right].
\]
\end{Theorem}

We close this section with some practical remarks. The first one is that the crucial part of Algorithm~\ref{gen:algo},
as far as valuation of the contingent claim is concerned, is composed of steps~1 and 2.  
The second one is that for real options and numerical simulations 
it is more convenient to work with undiscounted prices of the assets
and the contract. Thus, from now on we shall revert to actual prices and use a discounting factor of $\rho = \exp( r \Delta t)$
where $r$ is the risk-free rate.

%%%%%%%%%%%%%%%%%%  JPZ JULY 11th   %%%%%%%%%%%%%%%
If we are given a payment stream of cashflows, $c_t$ for $t=T_0,\cdots T_{F}\le \infty$, under the minimal martingale measure
$\hat{\PP}$ and discounting by the constant interest rate, the expected value $\VM_t$  is given by
$$
\VM_t = \widehat{\EE}\left[\sum_{s =t}^{T_F} c_s / \rho^{s-t}  \big|\F_t\right]  \mbox{ .}
$$
In this case, the generalization of Algorithm~\ref{gen:algo} is straightforward. 
Under the assumption that we are working in a Markovian setting such value becomes 
\begin{equation}\label{mark1}
 \VM_t = \widehat{\EE}\left[\sum_{s =t}^{T_F} c_s / \rho^{s-t}  \big| X_t = x \right]  \mbox{ .}
\end{equation}
% and it can be approximated by a variant of the HMC algorithm we shall describe in the next session. 
We shall now address the question of computing such conditional expectation from historical simulations. 
If we have a large number $N$ of simulations to the process $\left\{ X_t \right\}_{t=0,1,\cdots}$, we can approximate the
term on the R.H.S. of the local risk term 
$R^{\mathrm{loc}}_t$
by 
$$
R^{\mathrm{loc}}_t
\approx \frac{1}{N} \sum_{i=1}^{N}
\left(
\rho^{-1} V_{t+1}(X_{t+1}^i) - V_{t}(X_{t}^i) - \xi_{t+1}(X_{t}^i) \left[\rho^{-1}X_{t+1}^i - X_{t}^i \right] \right)^2\mbox{ .}
$$

The next step is to make the problem numerically tractable. But this, following the ideas of \cite{Longstaff2001} 
and \cite{potters2001hedged}, 
can be accomplished
by introducing a function basis for the unknown function $\xi_{t+1}(x)$ (respec. $V_{t}(x)$) and considering a finite element expansion. 
More precisely, let us write
% start mod by fm
$$V_t(x)=\sum_{a=1}^{b} \gamma_t^a K_a(x)$$ and 
$$\xi_{t+1}(x)=\sum_{a=1}^{b} \psi_{t+1}^a H_a(x) \mbox{ , }$$ 
% end mod by fm
where  $H_a$  (respec. $K_a$ ) forms a basis for the space of functions $\xi_{t+1}$ (respec. $V_t(x)$).
Then, one can substitute the minimization problem in Equation~(\ref{variation}) by the minimization:

% start mod by fm
\begin{equation} \label{numericmin}
\argmin_{\left\{\gamma_t^j, \psi_{t+1}^j \right\}_{j=1}^{b} }
\sum_{i=1}^{N}  \left[  \rho^{-1} V_{t+1}(X^i_{t+1}) - 
\sum_{a=1}^{b} \gamma_t^a K_a(X^i_t)- \right. \\
\left. \sum_{a=1}^{b} \psi_{t+1}^a H_a(X^i_t)\cdot(\rho^{-1} X^i_{t+1} -X^i_t) \right]^2 
\end{equation}
% end mod by fm

% \begin{Remark}
% In practice, the minimization of Equation~(\ref{variation}) is performed by using a number of different values of $X_t^i$ since
% usually the scenarios are generated as paths and not for a fixed value of $x$. 
% \end{Remark}
In other words, the expected value is computed by expanding the function in $L^2 (\Omega, \F_t, d \widehat{\PP})$ in
a suitable basis and truncating at an appropriate level. 
Needless to say, there are a number of relevant issues, ranging from conditions on the processes to approximation spaces. 
A more detailed analysis of the {\em non-Markovian} case and of such approximation spaces would take us too far afield. See for
example Section~1.3 of the work of \cite{Lipp2012}.

\subsection{The HMC Algorithm for Real Options}

We shall now present the proposed algorithm for the evaluation of the delay option of a project that could be started at any time between
say the time $T_0\ge 0$ and $T$. In financial terms, this consists of a Bermudian option that could be exercised at any time between 
$T_0$ and $T$. Obviously, it reduces to an American option if $T_0=0$ is the present time. In mathematical terms this corresponds
to a discrete version of a free boundary problem. 
We assume further that our
cash flows could come at any time till $T_F$. 
The main building block of our algorithm is the regression described in Equation~(\ref{numericmin}).

We assume we are given the following inputs:
\begin{itemize}
 \item A vector time series of traded assets $x_t^i$, for a period of times $t=T_0, \cdots, T$, and for the scenarios $i=1,\cdots, N$. 
 \item The corresponding cash flows associated to the different scenarios $c_t^i$ for $t=T_0, \cdots, T_{F}$, and 
 $i=1,\cdots, N$. Such cash flows would be produced by an oracle which takes into account the different traded asset values and
 the non-traded ones~\footnote{In principle, it could be also time dependent and even scenario dependent. Furthermore, 
 it can incorporate managerial views
 by emphasizing specific regions of the probability space.}.   
 \item A long term behavior for the project value or the cash flows (possibly under the different scenarios). 
 \item The exercise period of the optionality $T_0,\cdots,T$, where $0 \le T_0 < T \le T_{F}$. 
%  Usually, the initial investment time $T_0$ is
%  the present time $t=0$, but it is easy to envisage situations such that one would only be able to invest during specific periods. 
\end{itemize}

% {\bf Algorithmically}  
We now perform the following algorithm: 
\begin{Algorithm}{(HMC for Real Options)} \label{HMCAlg}
\begin{enumerate}
%  \item 
% Consider N paths $x^i_{t+1}$, $i=1,\cdots,N$
\item Initialize the project value $\VM_T(X^i_T) $ for the different scenarios $i=1,\cdots,N $ by using Equation~(\ref{mark1}) for $t=T_0\cdots T$.
% and different scenarios $i=1,\cdots,N $. 
\item Initialize for $t=T$ the payoff $\widehat{V}_{T}(X_T^i) = (\VM_T(X^i_T) - K)^{+}$ for the different scenarios
% and the auxiliary variable  $\widehat{V}_T^i=0$
$i=1,\cdots,N $. 
\item For $t=T-1, \cdots, T_0 $ do: 
\begin{enumerate}
\item
Define the functions: \\
% start mod by fm
$V_t(x):=\sum_{a=1}^{b} \gamma_t^a K_a(x)$ and $\xi_{t+1}(x):=\sum_{a=1}^{b} \psi_{t+1}^a H_a(x)$
\item Solve the quadratic minimization problem in terms of the coefficients $\gamma_t^a, \psi_{t+1}^a$:
\begin{equation*} %\label{numericmin}
\argmin_{\left\{\gamma_t^a, \psi_{t+1}^a \right\}_{a=1}^{b} }
\sum_{i=1}^{N} \left[ \rho^{-1} \widehat{V}_{t+1}(X^i_{t+1}) -
\sum_{a=1}^{b} \gamma_t^a K_a(X^i_t)-\right. \\
\left. \sum_{a=1}^{b} \psi_{t+1}^a H_a(X^i_t)\cdot(\rho^{-1} X^i_{t+1} -X^i_t) \right]^2
\end{equation*}
% end mod by fm

\item Define $\widehat{V}_t(X^i_t) := \max\{(\VM_t^i - K)^{+}, \widehat{V}_t(X^i_t) \}$.
% end mod by fm
\end{enumerate}
\item Output: The values of~ $\widehat{V}_{T_0}(x)$ for $x\in \left\{X_0^i \right\}_{i=1}^{N}$ and the points in the exercise region. 
\end{enumerate}
\end{Algorithm}

It $T_0=0$ we could continue the downward loop without the comparison and the computed values in $V_0$ 
would give an approximation for the option value  and the different 
scenarios~\footnote{Such different scenarios may reduce to a single point in case the initial scenario is known.} 
at the initial time $t=0$.

If we were working with the risk neutral simulations in a complete market, 
% the expected value of $(\rho^{-1} X^i_{t+1} -X^i_t)$ is zero and 
this algorithm reduces to a variant of the celebrated algorithm of 
\cite{Longstaff2001}.

\begin{Remark}\label{Felipe1}
In the actual implementation, the user may be interested in having access to the exercise region as well
as to more information about the suitability of investment by using different statistics. Thus, it may be interesting to 
refine the Item~3.c. of the algorithm as follows: 
% start mod fm
\begin{itemize} 
\item[3.c.]\quad Define $\widehat V_t(X^i_t) := \max\{(\VM_t(X_t^i) - K)^{+}, \widehat{V}_t(X^i_t) \}$ and store:
% Check for every $i=1,\cdots,N$ whether $\widehat{V}_t(X^i_t)\le (V_t^i - K)^{+}$. 
\begin{itemize}
 \item[ i.] $I_t:= \{i\in\{1,\cdots,N\}/ \widehat{V}_t(X^i_t) \leq (\VM_t(X_t^i) - K)^{+} \}$
 \item[ ii.]  $\nu_t:= \min\{ (\VM_t(X_t^i) - K) / i\in I_t \} $
%  \item $ P\left((V_t^i - K)\leq \nu_t \right)\approx \frac{|\{i\in\{1,\cdots,N\}/ V_t(X^i_t) \leq \nu_t \}|}{N} $
 \item[ iii.] $Pr_t:=P\left((\VM_t(X_t) - K)^{+}\leq \nu_t \right)
 \approx \#\{i\in\{1,\cdots,N\}/ (\VM_t(X_t^i) - K)^{+}\leq \nu_t \}\cdot N^{-1} $
 %$ P_t:=P\left((V_t^i - K)\leq \nu_t \right)\approx \#\{i\in\{1,\cdots,N\}/ (V_t^i - K)\leq \nu_t \}\cdot N^{-1} $
\end{itemize}
\end{itemize}
%end mod fm
The stored values of the points  $(t,\widehat{V}_t(X^i_t))$ for $i\in I_t$ correspond to an approximate description
of the exercise region.  

The quantity $\VM_t(X_t^i) - K$ will be called {\em intrinsic value of the investment option} in the sequel. It refers to the
best estimate of the stream of cash flows under the minimal martingale measure given the scenario $i$ minus the 
investment $K$.

The managerial usage of these statistics springs from the fact that, in many cases,  the stochastic generated cash flows inherit a corporate view of the market scenarios. As such, these statistics provide a subjective view on the investment scenarios that is appreciated by managers.

\end{Remark}

\paragraph{Implementation Notes}
The attentive reader will notice that the main bottle-neck of the whole procedure is precisely in the minimization of 3.(b). 
A fast and stable algorithm here would make the difference in practical applications. 
This minimization can be performed very efficiently
by using the QR algorithm to solve an overdetermined system of linear equations. See the text of \cite{GVL2013} for the numerical
analysis background.
The methodology can then be implemented (as we did) in a matlab-like environment with the standard Linpack packages. It can 
be easily ported to other popular programming languages such as R and Java.

The choice of the basis function is the subject of research by many authors even in the case of the
classical LSM algorithm of \cite{Longstaff2001}. We follow the  suggestion in the work of \cite{potters2001hedged} for the
one-dimensional case of taking the elements of the basis for hedge to be derivative of the ones for the option. We also 
take into account the suggested basis in \cite{GY2004}. 
In the 
multidimensional case we consider tensor products of the elements in the different dimensions.

\subsection{Remark on the continuous limits}

In the case of  data simulated or estimated  from a continuous model, we might consider realizations with  arbitrarily small 
time intervals and refined asset price grids. Then, a very natural question is whether  the  discrete algorithm has any form of limit as 
$\Delta t\searrow 0$. This problem then can be divided into two parts. First, the continuous limit of discrete time model.
Secondly, the numerical method to solve the limit case, its accuracy and efficiency.

Concerning the first issue, in the case of European options it is well established that the minimal martingale measure
of F\"olmer and Schweizer is associated to  Backward Stochastic Differential Equations (BSDEs). 
See for example \cite{EK:1997} for an
early account. 
In the work of 
\cite{PHAM2000} 
the main results of the theory of quadratic hedging in a 
general incomplete model of continuous trading with semi-martingale price process are reviewed. 
% The objective is to hedge contingent claims by using portfolio strategies. 
In particular, two types of criteria are studied: the mean-variance approach and 
the (local) risk-minimization, which is connected to the continuous limit of the approach considered here.
In the work of \cite{BS2004} the mean-variance hedging problem is
treated as a linear-quadratic stochastic control problem.
They show for continuous semi-martingales in a
general filtration that  the adjoint equations leads to BSDEs for the three coefficients of the quadratic value process.
 
Concerning the second issue, the use of regression-like Monte Carlo methods has received a lot of attention recently. 
See \cite{GLW2005, LGW2006, gobe:turk:13}
In particular, under appropriate conditions, the convergence of the HMC method can be proved and the error analysis has been performed
in \cite{gobe:turk:13}. Furthermore, in \cite{Lipp2012} the HMC method has been implemented to some exotic options and its numerical
aspects have  been studied.  In \cite{Gastel2013} the HMC method was implemented for actuarial problems. 

% In order to give an answer to this question, one has to consider  two steps: first it can be shown that, in the continuous case, 
% the value of the quadratic hedging portfolio satisfy a linear backward stochastic differential equation (BSDE)---cf.  \cite{EK:1997}. 
% The second step is to show that regression-like Monte Carlo methods are, in an appropriate sense,  discretization of such BSDEs, 
% as shown in  \cite{GLW2005}---see also \cite{LGW2006} for convergence rates associated to such discretization.

%The two ingredients  continuous time markets, the rescaling of the discrete times with spacing $1$ to the spacement
%$\Delta t$ is straightforward. Let us now suppose that we are working with a complete market driven by a diffusion process. 
%In this case,  if we fix the exercise time at say $[0,T_{E}]$ and let  $\Delta t \rightarrow 0$, 
%the convergence of this algorithm  is a consequence of general numerical methods to solve BSDEs 
%as discussed in the concluding section of the work of \cite{GLW2005}. See also \cite{LGW2006} for the convergence 
%rates associated to such methods.

\section{Examples and Case Studies}\label{examples}

We shall now exemplify the methodology proposed in the previous sections. The first set of examples will be purely illustrative ones 
aiming 
to exemplify the efficacy of the algorithm for option evaluation. They serve as validation and accuracy check for the codes. 
The second set comes from a large number of real data 
and practical evaluations. The examples take into account a large number of hedging energy commodities in the evaluation of a potential 
project in
the energy sector. Finally, we present an exploration on a fictitious example involving gas data (Henry Hub index) and a 
technology stock (Google). The 
project cash flows would be associated to the difference of (rescaled) values of such underlyings added to an uncorrelated and nonhedgeable
noise component. 

\subsection{Illustrative Theoretical Examples}
The first example concerns the running of the algorithm in the classical Black-Scholes market with simulated prices taken
in the historical measure. More precisely, we consider a European option expiring in 3 months 
with strike $K=100$, current asset price varying around the at-the-money value  $X(0)=100$, volatility  $\sigma=0.3$, and interest rate $r=0.05$.
% { MC: $\delta$time=1; \# basis (monomials)=3;  N=5000}
The number of basis elements (monomials $1$, $x$ and $x^2$) was $b=3$ and a total of $N=5000$ simulations in an arbitrary (fixed)
probability measure. 
\begin{figure}[t!]
% \begin{center}
\includegraphics[width=0.40\textwidth,angle=0]{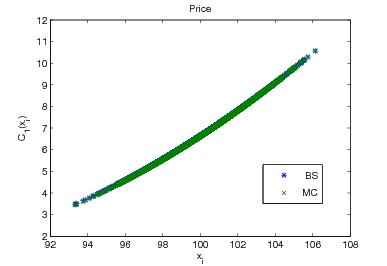}
% \hfil
\includegraphics[width=0.40\textwidth,angle=0]{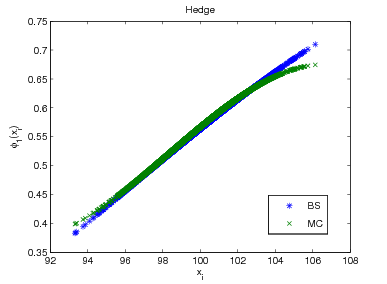}
% \end{center}
\caption{\label{ex001} The results of a comparison of the actual Black-Scholes formula price and the Hedged Monte Carlo algorithm
result. On the left we display the prices and on the right we display the hedge value.}
\end{figure}
% \FloatBarrier

Although this is a very simple text-book example, Figure~\ref{ex001}  conveys the fact that the results are pretty accurate even for such
a small number of simulations and small number of basis elements. 

% \begin{figure}
% \begin{center}
% \includegraphics[width=0.45\textwidth,angle=0]{hedgeBS.png}
% \end{center}
% \end{figure}
% { MC:  $\delta$time=1; \# basis (monomials)=3;  N=5000}

\begin{figure}[t!]
% \centering
\includegraphics[width=0.4\textwidth,angle=0]{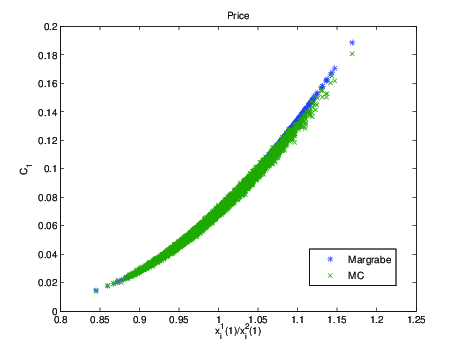}
\includegraphics[width=0.4\textwidth,angle=0]{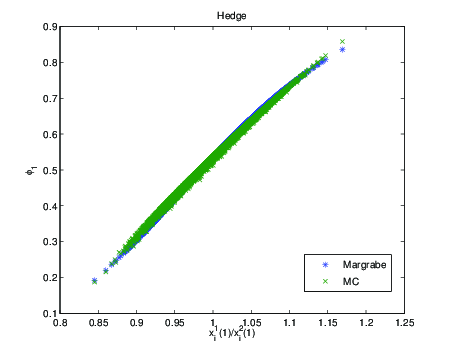}
\caption{\label{ex002} Results of the comparison between the HMC algorithm and the Margrabe formula. }
%A 65 days exchange option.
%On the left: price option as a function of the first underlying ($x^1_i(t)$) and the second one ($x^2_i(t)$) used as a {\it num\'eraire}, at time $t=1$. 
\end{figure}

In the second example we check the algorithm performance of the difference of two hedgeable assets $X_1$ and $X_2$. 
More precisely we consider a 65 days exchange option with payoff $(X_{1,T_{F}}-X_{2,T_{F}})^+$.
The variables $X_1$ and $X_2$ satisfy  geometrical Brownian motion dynamics
with 
$\sigma_1=0.3$, $\sigma_2=0.2$, and $r=0.05$. The analytical results are obtained using the Margrabe's formula.
% {\bf Options depending on various underlyings}
% { A 65 days exchange option with payoff $(x^1(T_{F})-x^2(T_{F}))^+$; $\sigma_1=0.3$, $\sigma_2=0.2$, $r=0.05.
In our setting this formula states that the fair price for the option is: $X_{1,0} N(d_1) - X_{2,0} N(d_2)$,
where $N$ denotes the cumulative distribution function for a normal distribution and 
$d_{1,2} = \left(\ln[X_{1,0}/X_{2,0}] \pm \sigma^2 T_{F}/2\right)/\sigma \sqrt{T_{F}}$, with $\sigma=\sqrt{0.3^2+0.2^2}$.
See \cite{musielarutkowski}. 
Here, we used two monomials and $N=10000$ simulations. 
The results are displayed in Figure~\ref{ex002}. 
% { MC:  $\delta$time=3; \# basis (monomials)=2;  N=10000, $m.t.<25$ sec}
% 
% 
% \begin{figure}
% \begin{center}
% \includegraphics[width=0.45\textwidth,angle=0]{HedgeMarg.png}
% \end{center}
% \end{figure}
% { MC:  $\delta$time=3; \# basis (monomials)=2;  N=10000, $m.t.<25$ sec}
% \FloatBarrier
\subsection{Practical Examples}
\paragraph{First Example}
An energy company considers the optionality of starting a new project that would last for $11$ years.
The project value $V_t$ is dependent on $12$ different underlyings. The option is exercizable every year during the first $5$ years. 
The company also has a trading desk that could be used for financial investment in some or all of such
different assets. 

The optionality was evaluated using several different sets of hedging assets. 
We now report on the results obtained with one hedging variable (in this example the Brent price) and 
considering $2000$ paths along $11$ years 
with a (continuously compounded annualized) interest rate $r=0.08$. 
We also computed examples with more hedging variables.
%The option is exercisable once at each year and the expiration  tim
% {\small Option evaluation using  {\bf all the hedging variables}}
% \begin{figure}
% \includegraphics[width=9.5cm,angle=0]{12Underlying.png}
% \end{figure}
% {\tiny $(V(t=0,S_1)-K)$ and $RO_{t=0}(S_1)$. $K=29.56$, $r=0.08$.
% %as a function of the Brent value.}
% }
%RO=109.7836mean, V-K=88.6754, K=29.5585
% \end{frame}
% e is $5$ years.
%The investment value was taken to be 25\% the average value of the future cash flow stream

\begin{figure}[h!]%\centering
% [b]
\sidecaption
\includegraphics[width=0.7\textwidth,angle=0]{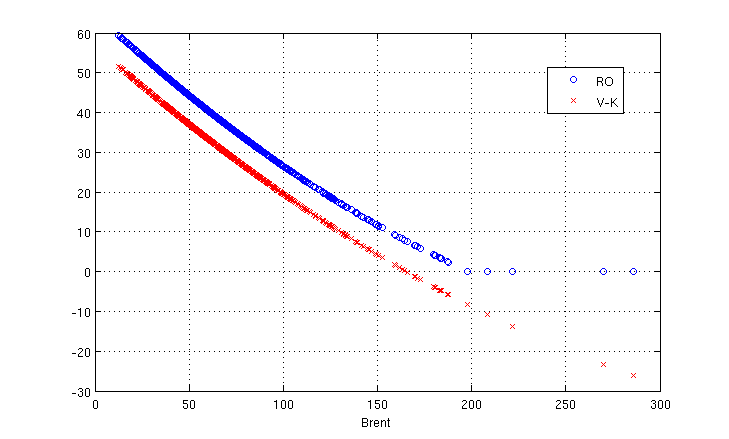}
\caption{Option evaluation using one hedging variable as a function of the Brent value. 
The difference between the project and the investment $(\VM_{t=0}(X_1)-K)$ is plotted
in (red) crosses while the optionality  $V_{t=0}(X_1)$ is plotted with (blue) circles. Here, the investment (strike) is $K=10.89$ and  
the risk free interest rate $r=0.08$.
 \label{example1}}
\end{figure}

In Figure~\ref{example1} we present the option evaluation using { one hedging variable}.
In this example the project works as a hedge towards low prices of the Brent. The fact that the intrinsic value of the project
is smaller than the optionality indicates that the company should wait to start the project.

\paragraph{Second Example}

\begin{figure}[h!]
 \centering
 \includegraphics[width=0.6\textwidth,angle=0]{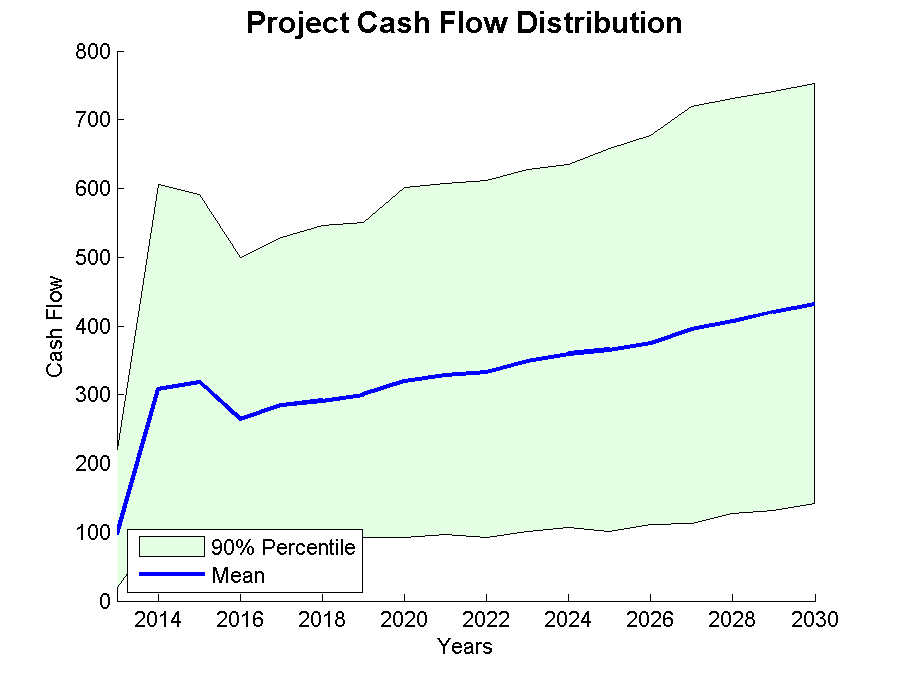}
 \caption{A description of the cash flow under the different scenarios. 
 The lower line corresponds to the 5\% quantile and top one to the 95\%. The marked region indicates
 the 90\% frequency region.
 \label{example2.1} 
}
\end{figure}

In this example we consider a project that would run for $15$ years, an investment of $1500$ monetary units and a yearly free interest rate
of $8.00\%$. The cash flows for this period are the results of an oracle that depends on a number of traded and non-tradable variables and
in turn are produced by means of
running different scenarios.  Some of their descriptive statistics is presented in Figure~\ref{example2.1}. 
%%%%%%%%%%%%%% FIG. 5 %%%%%%%%%%%%%%%%%%%%%%%%%%%%%%%%%%%

\begin{figure}[h!]
  \centering
  \includegraphics[width=0.6\textwidth,angle=0]{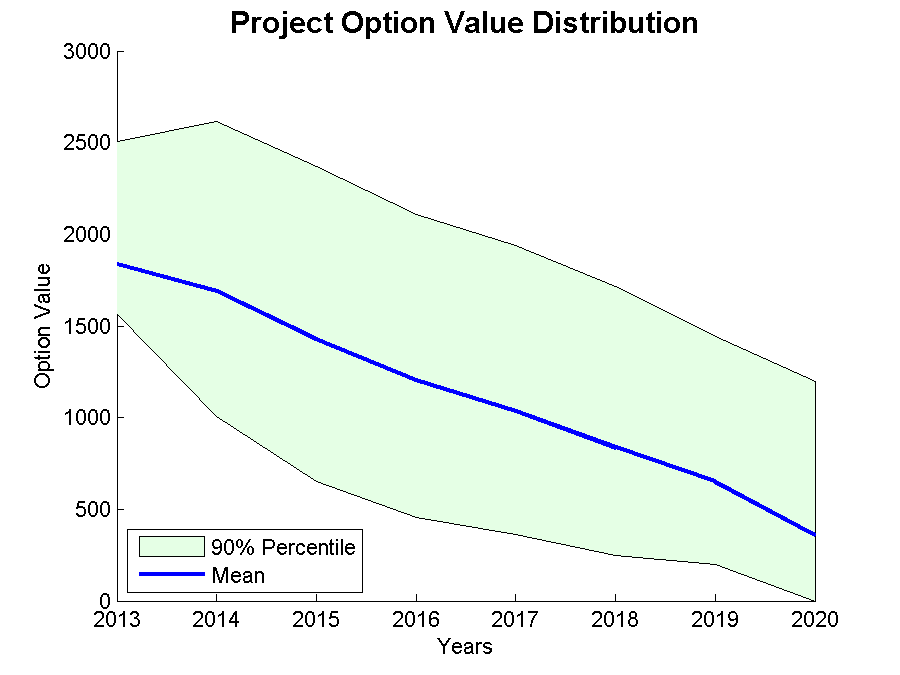}
  \caption{Value of the project optionality.  The lower line corresponds to the 5\% quantile and top one to the 95\%. The marked region indicates
 the 90\% frequency region. \label{example2.3} }
\end{figure}

\begin{figure}[h!]
 \centering
 \includegraphics[width=0.6\textwidth,angle=0]{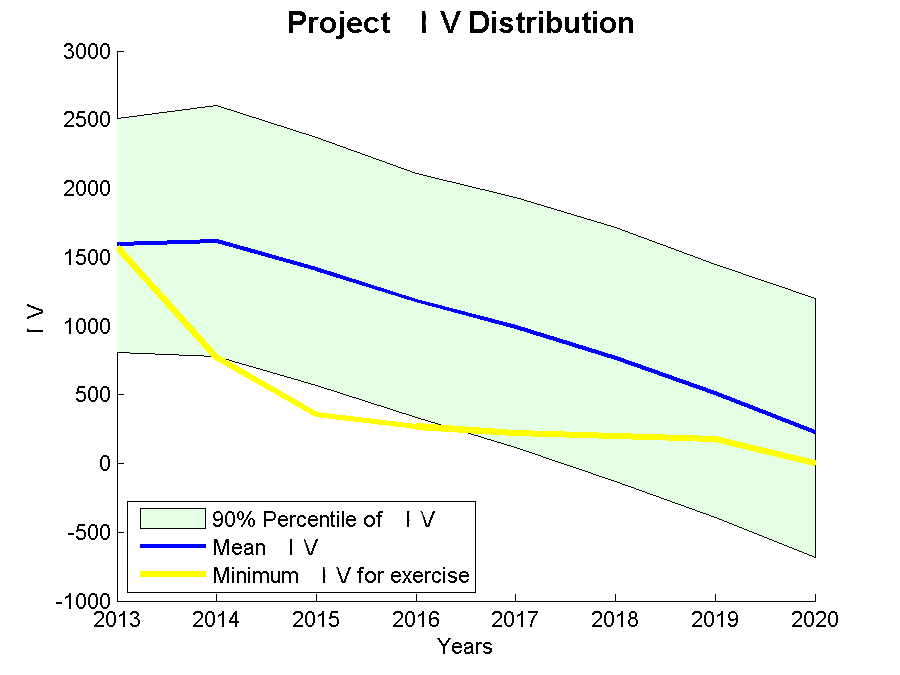}
 \caption{A description of the project \NPV (IV) statistics under the different scenarios and 
 the minimum value for exercise.
  The lower line corresponds to the 5\% quantile and top one to the 95\%. The marked region indicates
 the 90\% frequency region.
 \label{example2.4} 
}
\end{figure}

The % net present value %%%%%%%%%%%%%%%CHANGE  
intrinsic values of the optionality 
for the different times, including the  5\%,  and 95\% quantiles for the project value are 
presented in Figures~\ref{example2.3} and \ref{example2.4}.
By applying the Hedged Monte Carlo method we compute the value of the 
delay optionality considering $3$ hedging variables. %: \textbf{Brent, Maya} and \textbf{Marlim}. 
The project should be exercised if at a certain time and corresponding scenario the 
intrinsic project value is more than the delay optionality. 
%, i.e. its present value minues the 
% invested value, falls below the option value. 
This leads to a trigger curve that tells us for each scenario whether to invest or
not (Figure~\ref{example2.4}).
% \begin{frame}{Example from an Investiment Application (30 years)}{Net Present Value of the Project}

% \end{frame}

% \end{frame}
%*****************************************mod fm end ********************************************************************

\paragraph{Third Example}
Differently from the previous examples whereby the actual cash flows came from complex (black-box type) oracles, our present 
example concerns a fictitious project where the cash flows would come from a (fairly) simple mathematical function. 
It concerns an artificial potential 
investment on a gas propelled vehicle that could be used by an information technology company to gather geographical 
data and to use in their web-based advertisements. For simplicity we take the cash flow highly correlated to Google stock through 
the equation 
\begin{equation}\label{oracle}
c_t(X,\epsilon) = \mathsf{H}\left( a X_{1,t} - b X_{2,t}  - I + \epsilon_t \right) \mbox{ ,}
\end{equation}
where $X_1$ is the price of a Google stock, $X_2$ is Henry Hub (HH) gas index, $I$ is a fixed running cost, $\epsilon_t$ is a nonhedgeable
noise. The function $H$ in our example is defined as
$$
\mathsf{H}(x) = \left\{ \begin{array}{cc}
               0\mbox{ ,} & x \le 0  \mbox{ ,}\\
               x\mbox{ ,} & x \in (0,1) \mbox{ ,}\\
               1\mbox{ ,} & x \ge 0 \mbox{ .}
               \end{array}
\right.
$$
The rationale behind $\mathsf{H}$ is to simulate the saturation given by very large values of the stock and to clip the values below zero.

\begin{figure}[h!]
  \centering
  \includegraphics[width=0.9\textwidth,angle=0]{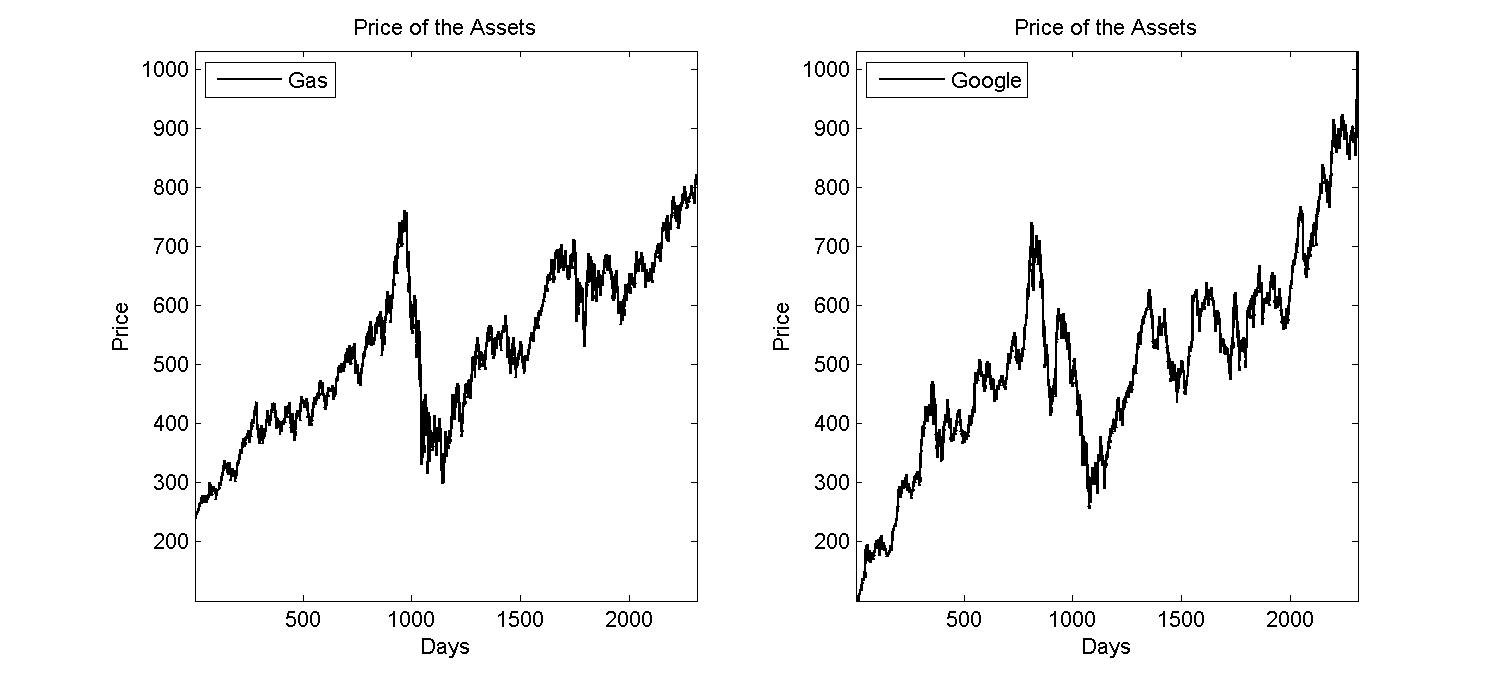}
  \caption{Time series for the assets between August 19th, 2004 and November 24th, 2013. \label{example3.1} }
\end{figure}

We performed the data collection using publicly available data downloaded by using public domain {\em R} 
software~\footnote{See for example \cite{RCite}.}.  The historical results  between August 19th, 2004 and November 24th, 2013
are displayed in Figure~\ref{example3.1}.
We calibrated the historical log-returns of the data with a GARCH(1,1) model, 
and then performed a principal component analysis of the bi-dimensional innovation time  
series. From that we generated the simulations of future scenarios. 

In this example we consider a project that would run for a maximum period of say $3$ years and the decisions could be performed monthly. 
The cash flows for this period are the results of
the oracle described in Equation~(\ref{oracle}) that depends on a value of Google and HH Gas. 
Finally we choose an investment of $INV=3.5$ and a risk-free interest rate of $8.00\%$.

% Using daily data of the asset between the 
% \textbf{ESCEREVER AS DATAS} we model scenarios using as GARCH(1,1) (\textbf{REFERENCIAS}) model.

\begin{figure}[h!]
  \centering
  \includegraphics[width=0.9\textwidth,angle=0]{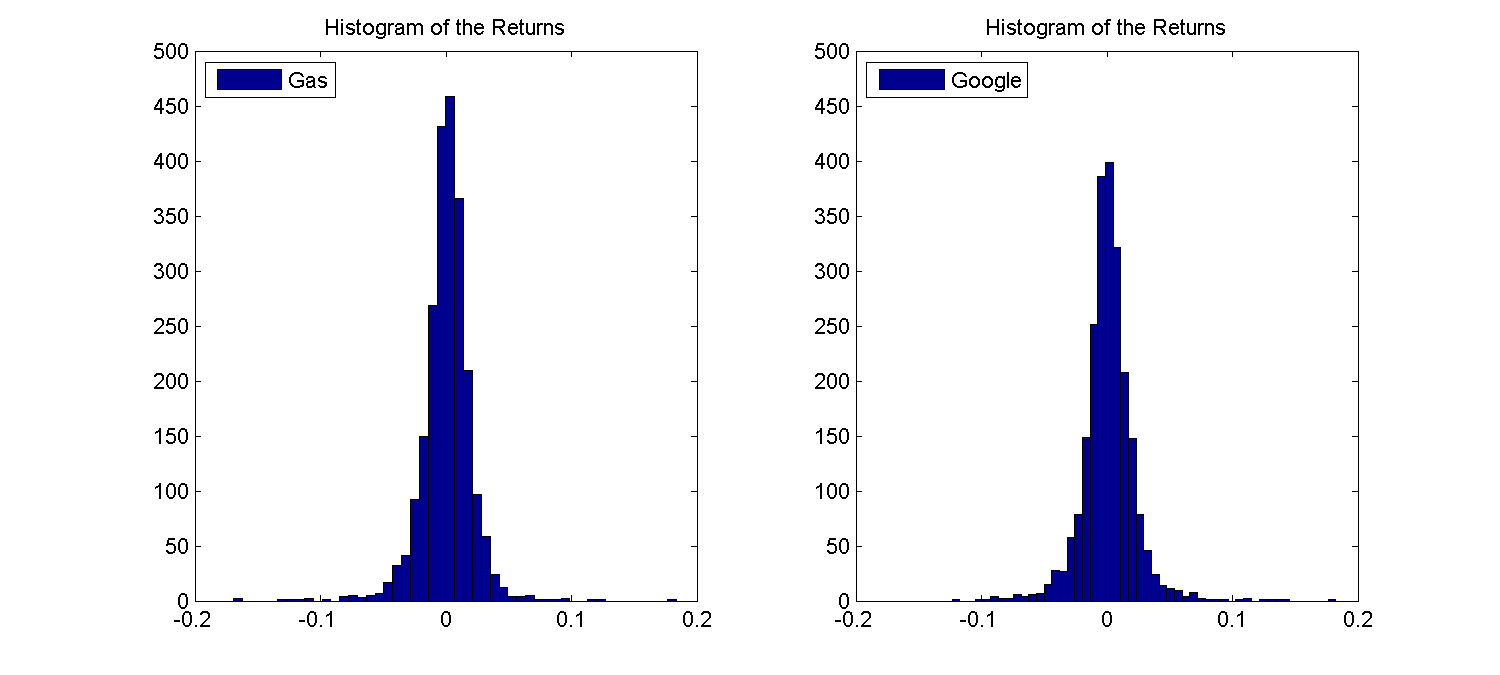}
  \caption{Histogram of the log returns for the assets between August 19th, 2004 and November 24th, 2013. \label{example3.2} }
\end{figure}

\begin{figure}[h!]
  \centering
  \includegraphics[width=0.9\textwidth,angle=0]{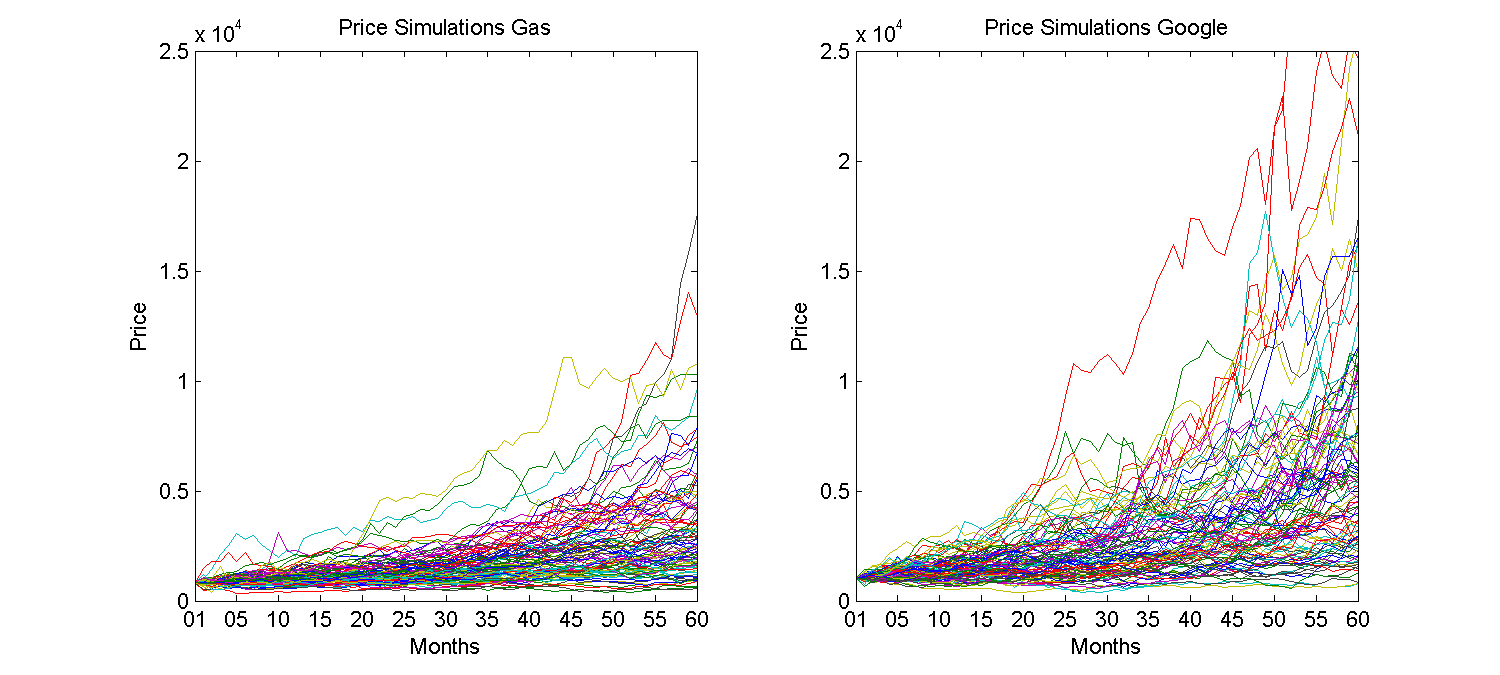}
  \caption{Asset simulations. \label{example3.3} }
\end{figure}

\begin{figure}[h!]
  \centering
\includegraphics[width=0.6\textwidth,angle=0]{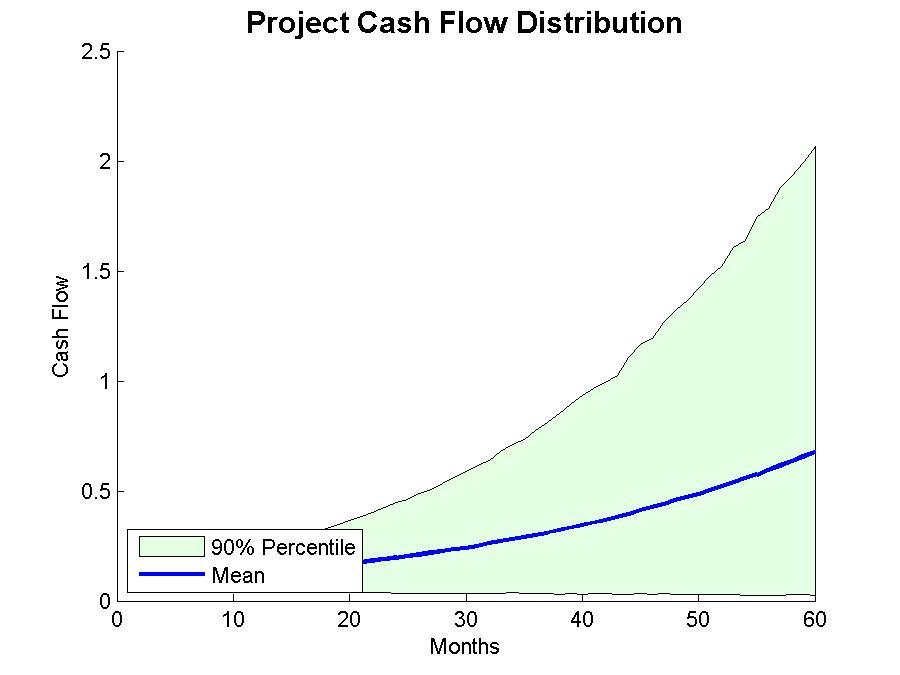}
  \caption{Cash flow simulations for the fictitious oracle described by Equation~(\ref{oracle}). Using the parameters value
   $a= 1.2895\times 10^{-4}$, $b=-5.3191\times 10^{-5}$, $I=0.05$, $\varepsilon_t\sim  \mathcal{N}(0, 0.005)$\label{example3.4} }
\end{figure}

In Figure~\ref{example3.3} we present some simulations of the assets, and in Figure~\ref{example3.4} a description of the 
simulations of the cash flows by showing their mean, their quantiles. 
% 
% \newpage

% \begin{equation*}
%  CF_t(X_1(t),X_2(t)):=T(aX_1(t)+bX_2(t)+c+\varepsilon_t)
% \end{equation*}
% Where
% \begin{equation*}
% \begin{split}
% T(x) & =0,\quad x \leq 0\\
% &=x,\quad x \in [0,1]\\
% &=1,\quad x \geq 1\\
% \end{split}
% \end{equation*}

% \begin{equation}
%  T(x):=\left{ 0, \quad x \leq 0 \\
% 	      1, \quad x \leq 0 \\
% 	      \right.
%  
% \end{equation}

\begin{figure}[h!]
  \centering
  \includegraphics[width=0.6\textwidth,angle=0]{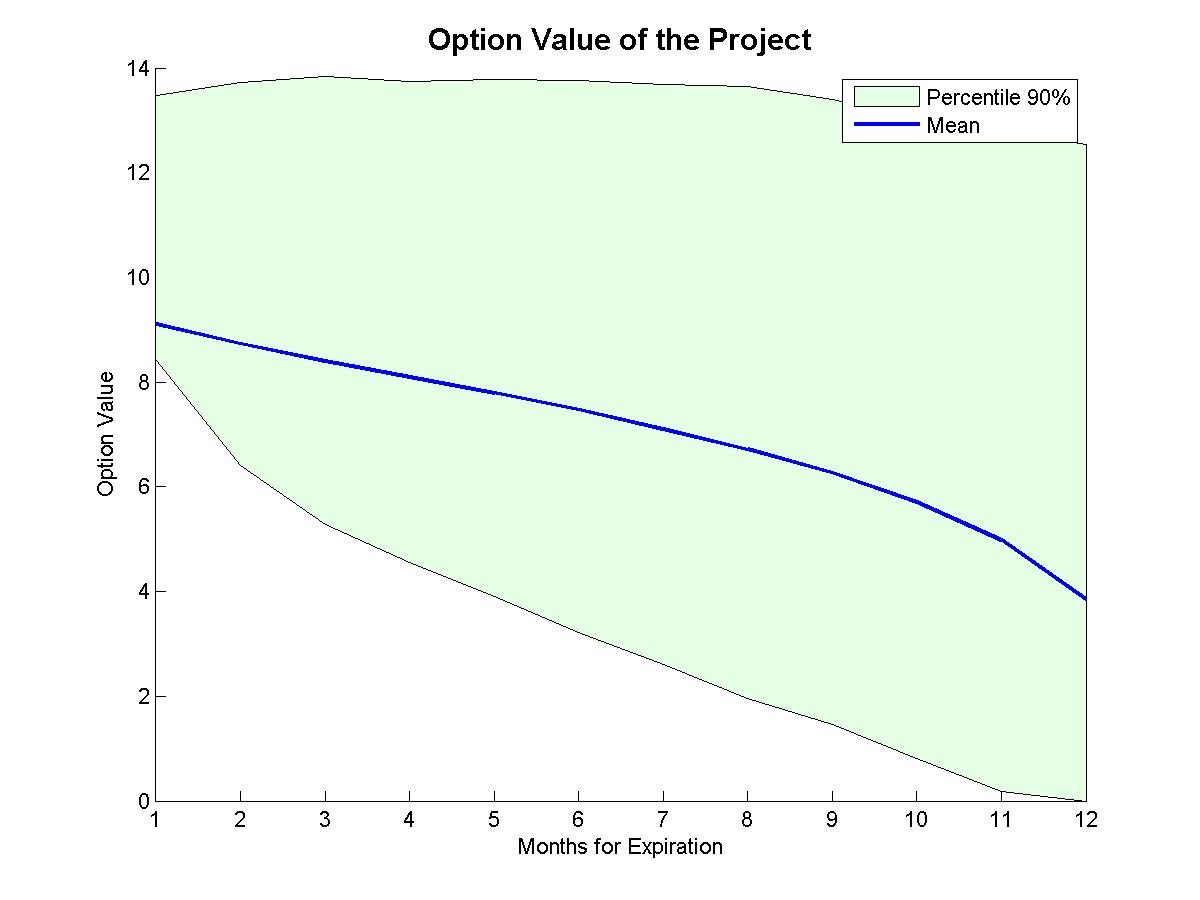}
  \caption{A description of the option value statistics under the different scenarios.
%   and the minimum value for exercise. 
  \label{example3.5} }
\end{figure}

\begin{figure}[h!]
  \centering
  \includegraphics[width=0.6\textwidth,angle=0]{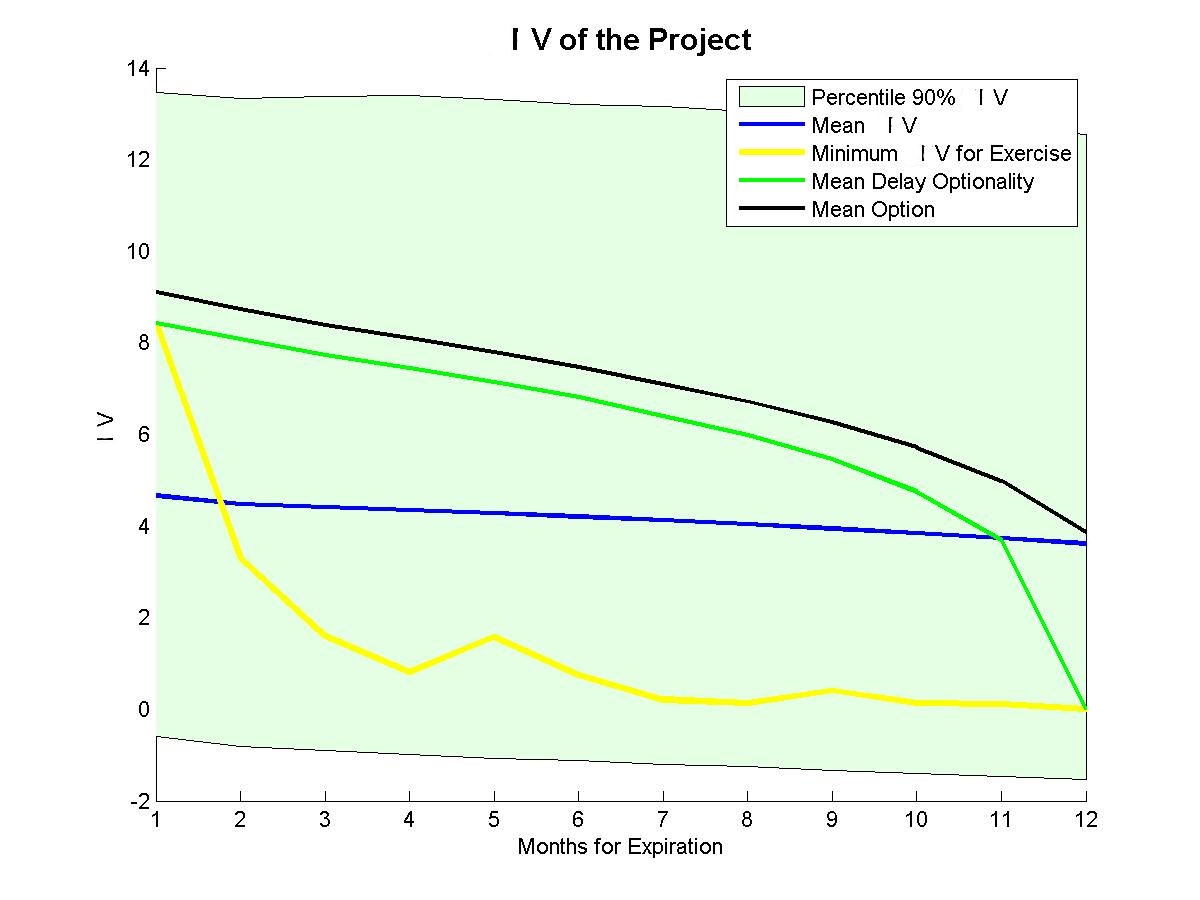}%CHANGE JPZ LAST MINUTE {Example1_NPV}
  \caption{A description of the project \NPV statistics under the different scenarios
  and the minimum value for 
  exercise. \label{example3.6} }
\end{figure}

The results in Figure~\ref{example3.6} show how the statistics of the values for the {\NPV} (defined as $\VM - I$)
relates to the curve of {minimum value of the \NPV for exercise} ($\nu_t$) that was calculated in the refined 
algorithm leading to Equation~(\ref{Felipe1}). 
% If the is small it indicates that is a good idea to do the investment at that time (in this case case t=10)
As the time varies between $t=1$ and $t=12$, the exercise curve crosses the average of the \NPVs for the different 
scenarios. %As it approaches zero, more scenarios would indicate exercise. 
The case of $\nu_t$ being smaller than the \NPV mean implies a small $Pr_t:=P(NPV<\nu_t)$. 
These small values of $Pr_t$ give a good suggestion of when to 
invest. But the decision to invest also has to involve the option value described in Figure~\ref{example3.5} and the expected \NPV value 
of Figure~\ref{example3.6}.

% \newpage
\section{Discussion and Conclusions}
\label{sec:conclude}

In this work we  addressed the problem of pricing  real options on  projects
that have their cash flow estimates based on an oracle prediction.
Such oracle is typically a combination of asset prices either used for
production or obtained as a result of the working project and non-traded
specific variables. They can also forecast prices or demand,  and they can
include managerial views or other non-tradable information that impacts 
the project value. These prices and variables may further be processed by an
optimization procedure, and this leads to the project cash flows. As discussed
in the Introduction, this appears naturally in many situations,
in particular for chemical or oil industries.

For such problems, we proposed a method that   is based on minimizing  the
tracking error variance of the hedge. This can be interpreted as assuming that
we are in an incomplete market and that the investor is naturally risk averse. 
In this context, 
this variance is a natural  risk measure for the investor. Under this
framework, we show  how to price real options using the method of \cite{potters2001hedged}.  
This lead to a set of consistent prices that reduces to that
of the Black-Scholes theory when the market is complete. The obtained price will depend on
the set of assets chosen for the hedge. This is natural since companies with
access to different markets and vulnerable to different scenarios can have
very different values for the same project. Theoretically, one could include
all hedging assets on a maximal set, but this is unfeasible from a practical
point of view.

Once more, we reinforce the idea that our simulations are all done in the historical measure where the calibration 
of the models take place. We could also have incorporated managerial views by emphasizing scenarios that would
be more likely due to management selective information. On the other extreme, even if the decision maker and the business at hand
had access to a completely correlated asset that could be used to hedge the project value,  
among the advantages of the present approach over a risk-neutral Monte Carlo evaluation we can mention: 
The reduction of variance of price estimation (for the same precision the number of paths can be up to 100 times smaller). This 
was already documented in the original work of \cite{potters2001hedged}. The estimation of the hedging strategy, residual risk (in the
form of the local variance), and possibly other risk measures (such as VaR and CVaR) at each time step. 

As explained in the conclusion of the work of \cite{G2011}, it is the time flexibility itself, more than the possibility of 
replication, that bears the extra value of an investment opportunity. Thus, the fact that we cannot replicate the project value should
not be the reason for not trying to quantify such extra value. The work of \cite{G2011} takes the point of view of utility functions
and indifference pricing. In contradistinction, here we took the point of view of minimizing risk as measured 
by the variance. A very natural follow up of the present work would be to compare the different approaches in the case of 
real world examples, such as the ones presented here. An exploration of the numerical issues related to the choice of
the projection basis would also be very welcome.

\begin{acknowledgement}

E.B. developed this work while visiting IMPA under the Cooperation Agreement between IMPA and Petrobras. 
MOS was partially supported by CNPq grant 308113/2012-8 and FAPERJ. 
JPZ was supported by CNPq grants 302161/2003-1 and 474085/2003-1 and by 
FAPERJ through the programs {\em Cientistas do Nosso Estado} and {\em Pensa Rio}.
All authors acknowledge the IMPA-PETROBRAS cooperation agreement. 

The authors would like to acknowledge  and thank a number of discussions with Fernando Aiube (PUC-RJ and Petrobras). 
We also thank Milene Mondek for the implementation of a number of preliminary examples of the HMC 
algorithm and Luca P. Mertens for help with the R software and the calibration procedure in the examples.

\end{acknowledgement}

\bibliographystyle{spbasic}
\bibliography{hmc}

\begin{thebibliography}{49}
\providecommand{\natexlab}[1]{#1}
\providecommand{\url}[1]{{#1}}
\providecommand{\urlprefix}{URL }
\expandafter\ifx\csname urlstyle\endcsname\relax
  \providecommand{\doi}[1]{DOI~\discretionary{}{}{}#1}\else
  \providecommand{\doi}{DOI~\discretionary{}{}{}\begingroup
  \urlstyle{rm}\Url}\fi
\providecommand{\eprint}[2][]{\url{#2}}

\bibitem[{Bobrovnytska and Schweizer(2004)}]{BS2004}
Bobrovnytska O, Schweizer M (2004) Mean-variance hedging and stochastic
  control: beyond the {B}rownian setting. IEEE Trans Automat Control
  49(3):396--408, \doi{10.1109/TAC.2004.824468},
  \urlprefix\url{http://dx.doi.org/10.1109/TAC.2004.824468}

\bibitem[{Borison(2005)}]{Borison2005}
Borison A (2005) Real options analysis: Where are the emperor's clothes?
  Journal of Applied Corporate Finance 17(2):17--31,
  \urlprefix\url{http://dx.doi.org/10.1111/j.1745-6622.2005.00029.x}

\bibitem[{Brennan and Schwartz(1985)}]{BrSc85}
Brennan MJ, Schwartz ES (1985) {Evaluating Natural Resource Investments}. The
  Journal of Business 58(2):pp. 135--157,
  \urlprefix\url{http://www.jstor.org/stable/2352967}

\bibitem[{Chen et~al(2007)Chen, Chiang, and Chien}]{CCC2007}
Chen CC, Chiang YS, Chien CF (2007) Real option analysis for capacity
  investment planning for semiconductor manufacturing. In: Semiconductor
  Manufacturing, 2007. ISSM 2007. International Symposium on, pp 1--3,
  \doi{10.1109/ISSM.2007.4446823}

\bibitem[{Choulli and Stricker(1996)}]{CS96Watanabe}
Choulli T, Stricker C (1996) Deux applications de la d\'ecomposition de
  {G}altchouk-{K}unita-{W}atanabe. In: S\'eminaire de {P}robabilit\'es, {XXX},
  Lecture Notes in Math., vol 1626, Springer, Berlin, pp 12--23,
  \doi{10.1007/BFb0094638},
  \urlprefix\url{http://dx.doi.org/10.1007/BFb0094638}

\bibitem[{Copeland and Antikarov(2001)}]{Copeland2001}
Copeland T, Antikarov V (2001) {Real Options: A Practitioner's Guide}. W. W.
  Norton and Company

\bibitem[{Copeland and Tufano(2004)}]{copeland2004real}
Copeland T, Tufano P (2004) {A real-world way to manage real options}. Harvard
  business review 82(3):90--99

\bibitem[{Dixit(1989)}]{DIXIT1989}
Dixit A (1989) {Entry And Exit Decisions Under Uncertainty}. Journal Of
  Political Economy 97(3):620--638

\bibitem[{Dixit and Pindyck(1994)}]{dixitpindyck94}
Dixit A, Pindyck R (1994) {Investment under Uncertainty}. Princeton University
  Press

\bibitem[{El~Karoui et~al(1997)El~Karoui, Peng, and Quenez}]{EK:1997}
El~Karoui N, Peng S, Quenez MC (1997) Backward stochastic differential
  equations in finance. Mathematical Finance 7(1):1--71,
  \doi{10.1111/1467-9965.00022},
  \urlprefix\url{http://dx.doi.org/10.1111/1467-9965.00022}

\bibitem[{F{\"o}llmer and Schied(2004)}]{Follmer:Schied:2004}
F{\"o}llmer H, Schied A (2004) Stochastic finance, volume 27 of de gruyter
  studies in mathematics

\bibitem[{Gastel(2013)}]{Gastel2013}
Gastel LV (2013) {Risk Beyond the Hedge: Options and Guarantees Embedded in
  Life Insurance Products in Incomplete Markets}. Master's thesis, University
  of Amsterdam

\bibitem[{Glasserman and Yu(2004)}]{GY2004}
Glasserman P, Yu B (2004) Number of paths versus number of basis functions in
  {A}merican option pricing. Ann Appl Probab 14(4):2090--2119

\bibitem[{Gobet and Turkedjiev(2013)}]{gobe:turk:13}
Gobet E, Turkedjiev P (2013) Linear regression {MDP} scheme for discrete
  backward stochastic differential equations under general conditions. to
  appear: Mathematics of Computation

\bibitem[{Gobet et~al(2005)Gobet, Lemor, and Warin}]{GLW2005}
Gobet E, Lemor JP, Warin X (2005) A regression-based {M}onte {C}arlo method to
  solve backward stochastic differential equations. Ann Appl Probab
  15(3):2172--2202, \doi{10.1214/105051605000000412},
  \urlprefix\url{http://dx.doi.org/10.1214/105051605000000412}

\bibitem[{Golub and Van~Loan(2013)}]{GVL2013}
Golub GH, Van~Loan CF (2013) Matrix computations, 4th edn. Johns Hopkins
  Studies in the Mathematical Sciences, Johns Hopkins University Press,
  Baltimore, MD

\bibitem[{Grasselli and Hurd(2004)}]{gh2004}
Grasselli M, Hurd T (2004) {A {M}onte Carlo method for exponential hedging of
  contingent claims}. Unpublished preprint

\bibitem[{Grasselli(2011)}]{G2011}
Grasselli MR (2011) {Getting Real with Real Options: A Utility--Based Approach
  for Finite--Time Investment in Incomplete Markets}. Journal of Business
  Finance \& Accounting 38(5-6):740--764,
  \doi{10.1111/j.1468-5957.2010.02232.x},
  \urlprefix\url{http://dx.doi.org/10.1111/j.1468-5957.2010.02232.x}

\bibitem[{Grasselli and Hurd(2007)}]{mr2349306}
Grasselli MR, Hurd TR (2007) {Indifference pricing and hedging for volatility
  derivatives}. Appl Math Finance 14(4):303--317,
  \doi{10.1080/13527260600963851},
  \urlprefix\url{http://dx.doi.org/10.1080/13527260600963851}

\bibitem[{Henderson and Hobson(2002)}]{HH2002}
Henderson V, Hobson DG (2002) Real options with constant relative risk
  aversion. Journal of Economic Dynamics and Control 27(2):329--355

\bibitem[{Hubalek and Schachermayer(2001)}]{HS2001}
Hubalek F, Schachermayer W (2001) {The limitations of no-arbitrage arguments
  for real options}. Int J Theor Appl Finance 2(4):361--373

\bibitem[{Ingersoll and Ross(1992)}]{INGERSOLL1992}
Ingersoll JE, Ross SA (1992) {Waiting To Invest - Investment And Uncertainty}.
  Journal Of Business 65(1):1--29

\bibitem[{Jaimungal and Lawryshyn(2011)}]{Jaimungal2011}
Jaimungal S, Lawryshyn Y (2011) {Incorporating Managerial Information into Real
  Option Valuation}. SSRN eLibrary

\bibitem[{Lemor et~al(2006)Lemor, Gobet, and Warin}]{LGW2006}
Lemor JP, Gobet E, Warin X (2006) Rate of convergence of an empirical
  regression method for solving generalized backward stochastic differential
  equations. Bernoulli 12(5):889--916, \doi{10.3150/bj/1161614951},
  \urlprefix\url{http://dx.doi.org/10.3150/bj/1161614951}

\bibitem[{Lipp(2012)}]{Lipp2012}
Lipp T (2012) {Numerical Methods for Optimization in Finance: Optimized Hedges
  for Options and Optmized Options for Hedging}. PhD thesis, Augsburg
  University and Pierre and Marie Curie University

\bibitem[{Longstaff and Schwartz(2001)}]{Longstaff2001}
Longstaff FA, Schwartz ES (2001) {Valuing American options by simulation: A
  simple least-squares approach}. Review of Financial Studies 14(1):113--147

\bibitem[{Mathews et~al(2007)Mathews, Datar, and Johnson}]{dm1}
Mathews S, Datar V, Johnson B (2007) {A practical method for valuing real
  options: The Boeing approach}. Journal of Applied Corporate Finance
  19(2):95--104

\bibitem[{McDonald and Siegel(1986)}]{MCDONALD1986}
McDonald R, Siegel D (1986) {The Value Of Waiting To Invest}. Quarterly Journal
  Of Economics 101(4):707--727

\bibitem[{Mittal(2004)}]{M2004}
Mittal G (2004) {Real Options Approach to Capacity Planning Under Uncertainty}.
  Master's thesis, Massachusetts Institute of Technology

\bibitem[{Moro et~al(1998)Moro, Zanin, and Pinto}]{MZP98}
Moro L, Zanin A, Pinto J (1998) A planning model for refinery diesel
  production. Computers and Chemical Engineering 22(1):1039--‐1042

\bibitem[{Musiela and Rutkowski(1997)}]{musielarutkowski}
Musiela M, Rutkowski M (1997) {Martingale methods in financial modeling}.
  Springer-Verlag, New York, NY

\bibitem[{Myers(1977)}]{MYERS1977}
Myers SC (1977) {Determinants Of Corporate Borrowing}. Journal Of Financial
  Economics 5(2):147--175

\bibitem[{Novaes and Souza(2005)}]{NS2005}
Novaes AGN, Souza JC (2005) A real options approach to a classical capacity
  expansion problem. Pesquisa Operacional 25(2):159--181

\bibitem[{Oldenburg et~al(2007)Oldenburg, Schlegel, Ulrich, Hong, Krepinsky,
  Grossmann, Polt, Terhorst, and Snoeck}]{Oetal2007}
Oldenburg J, Schlegel M, Ulrich J, Hong TL, Krepinsky B, Grossmann G, Polt A,
  Terhorst H, Snoeck JW (2007) A method for quick evaluation of stepwise plant
  expansion scenarios in the chemical industry. In: Plesu V, Agachi P (eds)
  17th European Symposium on Computer Aided Process Engineering, Elsevier B.V.,
  p~2

\bibitem[{Paddock et~al(1988)Paddock, Siegel, and Smith}]{PADDOCK1988}
Paddock JL, Siegel DR, Smith JL (1988) {Option Valuation Of Claims On Real
  Assets - The Case Of Offshore Petroleum Leases}. Quarterly Journal Of
  Economics 103(3):479--508

\bibitem[{Papageorgiou(2009)}]{P2009}
Papageorgiou LG (2009) Supply chain optimization for the process industries:
  Advances and opportunties. Computers and Chemical Engineering
  33(1):11,931--‐1938

\bibitem[{Pham(2000)}]{PHAM2000}
Pham H (2000) On quadratic hedging in continuous time. Mathematical Methods of
  Operations Research 51(2):315--339, \doi{10.1007/s001860050091},
  \urlprefix\url{http://dx.doi.org/10.1007/s001860050091}

\bibitem[{Pindyck(1991)}]{PINDYCK1991}
Pindyck RS (1991) {Irreversibility, Uncertainty, And Investment}. Journal Of
  Economic Literature 29(3):1110--1148

\bibitem[{Potters et~al(2001)Potters, Bouchaud, and
  Sestovic}]{potters2001hedged}
Potters M, Bouchaud J, Sestovic D (2001) {Hedged {M}onte-Carlo: low variance
  derivative pricing with objective probabilities}. Physica A: Statistical
  Mechanics and its Applications 289(3-4):517--525

\bibitem[{Primbs and Yamada(2008)}]{PY2008}
Primbs JA, Yamada Y (2008) A new computational tool for analysing dynamic
  hedging under transaction costs. Quantitative Finance 8(4):405--413

\bibitem[{{R Core Team}(2013)}]{RCite}
{R Core Team} (2013) R: A Language and Environment for Statistical Computing. R
  Foundation for Statistical Computing, Vienna, Austria,
  \urlprefix\url{http://www.R-project.org/}

\bibitem[{Sahinidis et~al(1989)Sahinidis, Grossmann, Fornari, and
  Chathrathi}]{SGFC1989}
Sahinidis N, Grossmann I, Fornari R, Chathrathi M (1989) Optimization model for
  long range planning in the chemical industry. Computers and Chemical
  Engineering 13:1049--‐1063

\bibitem[{Schweizer(2008)}]{Schw2008}
Schweizer M (2008) {Local Risk-Minimization for Multidimensional Assets and
  Payment Streams}. Banach Center Publications 83:213--229

\bibitem[{Schweizer and F\"{o}llmer(1988)}]{SF1988}
Schweizer M, F\"{o}llmer H (1988) {Hedging by Sequential Regression: an
  Introduction to the Mathematics of Option Trading}. ASTIN Bulletin
  18(2):147--160

\bibitem[{Shapiro(2009)}]{S2009}
Shapiro JF (2009) Challenges of strategic supply chain planning and modeling.
  Computers and Chemical Engineering 28:855--861

\bibitem[{Titman(1985)}]{TITMAN1985}
Titman S (1985) {Urban Land Prices Under Uncertainty}. American Economic Review
  75(3):505--514

\bibitem[{Tourinho(1979)}]{Touri1979}
Tourinho O (1979) {The Option Value of Reserves of Natural Resources}. Working
  Paper 94, University of California, Berkeley

\bibitem[{Trigeorgis(1999)}]{Trigeorgis1999}
Trigeorgis L (1999) {Real Options: Managerial Flexibility and Strategy in
  Resource Allocation}. The MIT Press

\bibitem[{Trigeorgis and Mason(1987)}]{TG1987}
Trigeorgis L, Mason SP (1987) {Valuing managerial flexibility}. Midland
  Corporate Finance Journal 5(1):14--21

\end{thebibliography}
\end{document}